\documentclass[aps,prb,reprint,groupedaddress,showpacs,amsmath,amssymb,superscriptaddress]{revtex4-1}
\usepackage{graphicx}
\usepackage{caption}
\usepackage{dcolumn}% Align table columns on decimal point
\usepackage{bm}% bold math
\usepackage{float}
\usepackage{url}
\usepackage{amssymb}

\usepackage{subfigure}
\usepackage{calc}
\usepackage{latexsym}
\usepackage{epsf}
\usepackage{epsfig}
\usepackage{notoccite}
\usepackage{euscript}
\usepackage{hyperref}
\usepackage{xcolor}
\usepackage{sidecap}
\usepackage[english]{babel}
\usepackage[autostyle, english = american]{csquotes}
\MakeOuterQuote{"}
\usepackage{soul} %For highlighting text with \hl{text}
\usepackage{gensymb}%Allow for symbols such as the degrees symbol to be used
\usepackage{textcomp}%Also allows for symbols such as the degrees symbol to be used, but directly within text.

\begin{document}

 %Use the \preprint command to place your local institutional report
 %number in the upper righthand corner of the title page in preprint mode.
 %Multiple \preprint commands are allowed.
 %Use the 'preprintnumbers' class option to override journal defaults
 %to display numbers if necessary
%\preprint{}

%Title of paper
\title{Modification of Spin Ice Physics in Ho$_2$Ti$_2$O$_7$ Thin Films}

% repeat the \author .. \affiliation  etc. as needed
 %\email, \thanks, \homepage, \altaffiliation all apply to the current
 %author. Explanatory text should go in the []'s, actual e-mail
% address or url should go in the {}'s for \email and \homepage.
% Please use the appropriate macro foreach each type of information

 %\affiliation command applies to all authors since the last
% \affiliation command. The \affiliation command should follow the
 %other information
 %\affiliation can be followed by \email, \homepage, \thanks as well.
\author{Kevin Barry}
 \affiliation{Department of Physics, FSU, Tallahassee, FL 32310, USA}
 \affiliation{National High Magnetic Field Laboratory, FSU, Tallahassee, FL 32310, USA}
 
 \author{Biwen Zhang}
 \affiliation{Department of Physics, FSU, Tallahassee, FL 32310, USA}
 \affiliation{National High Magnetic Field Laboratory, FSU, Tallahassee, FL 32310, USA}
 
 \author{Naween Anand}
 \affiliation{National High Magnetic Field Laboratory, FSU, Tallahassee, FL 32310, USA}
 
 \author{Yan  Xin}
  \affiliation{National High Magnetic Field Laboratory, FSU, Tallahassee, FL 32310, USA}

\author{Arturas Vailionis}
\affiliation{Stanford Nano Shared Facilities, Stanford University, Stanford, CA 94305, USA}
 
\author{Jennifer Neu}
 \affiliation{Department of Physics, FSU, Tallahassee, FL 32310, USA}
 \affiliation{National High Magnetic Field Laboratory, FSU, Tallahassee, FL 32310, USA}
 
\author{Colin Heikes}
\affiliation{NIST Center for Neutron Research, National Institute of Standards and Technology, Gaithersburg, MD 20899, USA}
 
\author{Charis Cochran}
 \affiliation{National High Magnetic Field Laboratory, FSU, Tallahassee, FL 32310, USA}

\author{Haidong Zhou}
 \affiliation{National High Magnetic Field Laboratory, FSU, Tallahassee, FL 32310, USA}
\affiliation{University of Tennessee, Knoxville, TN 37996, USA}

\author{Y. Qiu}
\affiliation{NIST Center for Neutron Research, National Institute of Standards and Technology, Gaithersburg, MD 20899, USA}

\author{William Ratcliff}
\affiliation{NIST Center for Neutron Research, National Institute of Standards and Technology, Gaithersburg, MD 20899, USA}

\author{Theo Siegrist}
\affiliation{National High Magnetic Field Laboratory, FSU, Tallahassee, FL 32310, USA}
\affiliation{Department of Chemical and Biomedical Engineering, FAMU-FSU College of Engineering, FL 32310, USA}

\author{Christianne Beekman}
 \affiliation{Department of Physics, FSU, Tallahassee, FL 32310, USA}
 \affiliation{National High Magnetic Field Laboratory, FSU, Tallahassee, FL 32310, USA}
\date{\today}
\begin{abstract}
\noindent We present an extensive study on the effect of substrate orientation, strain, stoichiometry and defects on spin ice physics in Ho$_2$Ti$_2$O$_7$ thin films grown onto yttria-stabilized-zirconia substrates. We find that growth in different orientations produces different strain states in the films. All films exhibit similar c-axis lattice parameters for their relaxed portions, which are consistently larger than the bulk value of 10.10 \AA. Transmission electron microscopy reveals anti-site disorder and growth defects to be present in the films, but stuffing is not observed. The amount of disorder depends on the growth orientation, with the (110) film showing the least. Magnetization measurements at 1.8 K show the expected magnetic anisotropy and saturation magnetization values associated with a spin ice for all orientations; shape anisotropy is apparent when comparing in and out-of-plane directions. Significantly, only the (110) oriented films display the hallmark spin ice plateau state in magnetization, albeit less well-defined compared to the plateau observed in a single crystal. Neutron scattering maps on the more disordered (111) oriented films show the Q=0 phase previously observed in bulk materials, but the Q=X phase giving the plateau state remains elusive. We conclude that the spin ice physics in thin films is modified by defects and strain, leading to a reduction in the temperature at which correlations drive the system into the spin ice state.
\end{abstract}
%\pacs{78.20.-e,78.20.Ci,78.40.Fy}
\maketitle

\section{INTRODUCTION}

The quest for novel quantum phases that show collective degrees of freedom and fractionalized excitations is one of the central themes in condensed matter physics. There has been a strong focus recently on geometrically frustrated magnets, which have been shown to host a number of cooperative spin states \cite{Ramirez}, such as spin ice and spin liquid states, in which fractionalized excitations (topological defects that behave like magnetic monopoles) have been experimentally realized \cite{gingras2014, Castelnovo, Jaubert, JaubertLin}. Of particular interest are rare earth pyrochlores that belong to the spin ice family, such as Ho$_{2}$Ti$_{2}$O$_{7}$, which display severe frustration that arises from a combination of the lattice geometry, and the sign of the nearest neighbor magnetic interactions resulting in local ice rules, and a macroscopically degenerate ground state.  One important factor determining the ground state selection in these systems is the single ion anisotropy \cite{cao,Tomasello}.  

\begin{figure}[h]
\centering
\includegraphics[width= 3.5 in,height= 3.5 in,keepaspectratio]{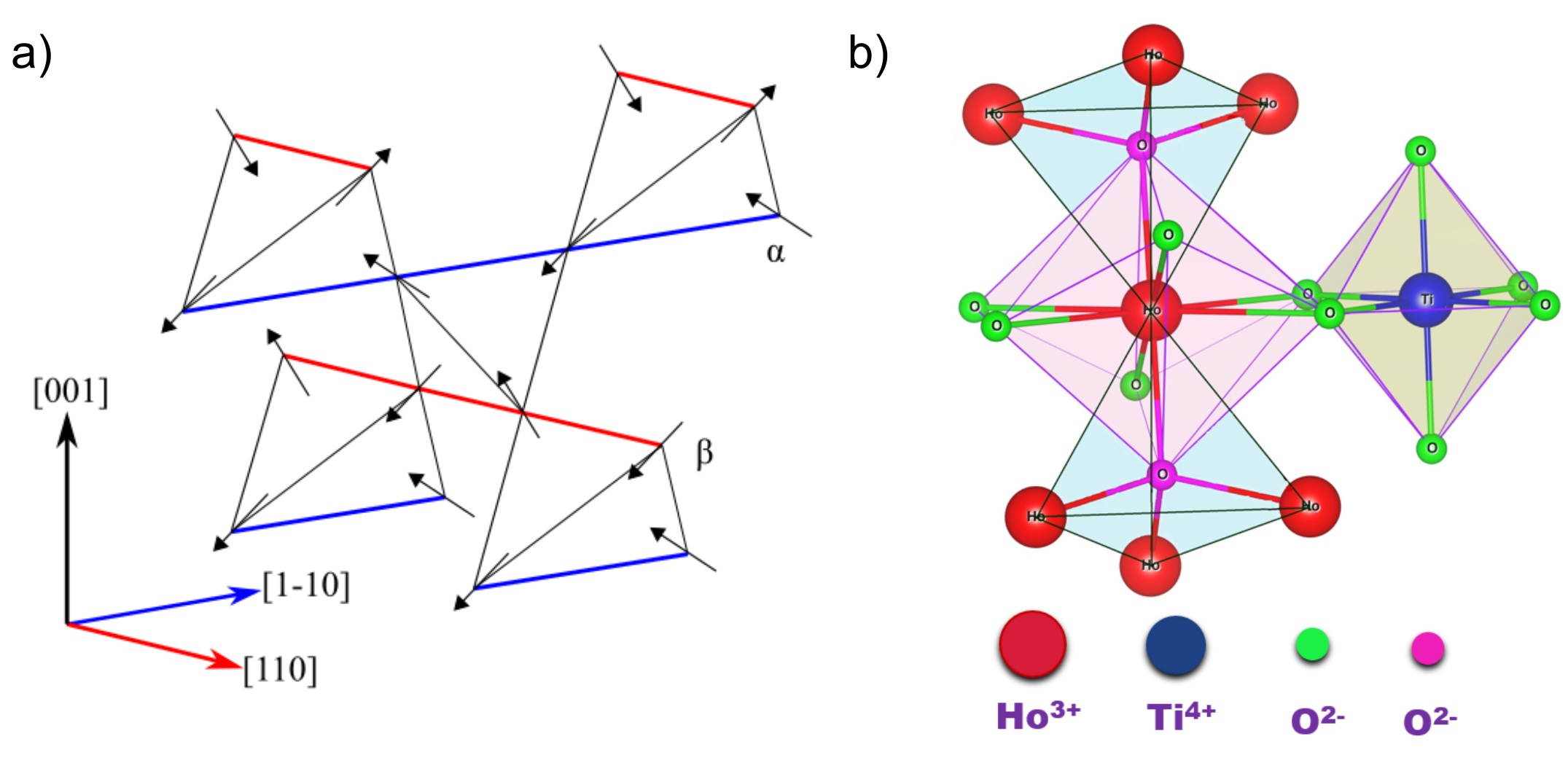}
\caption{\label{CS12}(Color online) a) Corner sharing tetrahedral network of Ho atoms in Ho$_{2}$Ti$_{2}$O$_{7}$ pyrochlore structure showing a two-in and two-out magnetic configuration. The $\alpha$ and $\beta$ chains are indicated in blue and red along with their respective crystallographic directions (similar to ref. \cite{clancy}). b) Anisotropic crystal field around Ho atom leading to two inequivalent oxygen sites (O$_1$ in a ring around the Ho and O$_2$ as two apical oxygen atoms along the \textless111\textgreater) and strong axial symmetry along \textless111\textgreater \ local axis. (red atoms: Ho; blue atoms: Ti; green atoms: O$_1$; violet atoms: O$_2$)}
\end{figure}

Ho$_{2}$Ti$_{2}$O$_{7}$ has a cubic structure (Fd$\bar{3}$m space group) and can be envisioned as Ho$^{3+}$ ions (each with a magnetic moment of $\sim$10 $\mu_B$) residing on a lattice of cornersharing tetrahedra (see Fig. \ref{CS12} a)) \cite{gardner}. In Ho$_{2}$Ti$_{2}$O$_{7}$ the crystal field doublet ground state results in an Ising-like easy axis anisotropy (see Fig. \ref{CS12} b)) with all the spins pointing along the local $<111>$ axes, i.e. either in or out of the tetrahedra \cite{gardner,clancy}. If the system were dominated by antiferromagnetic nearest neighbor exchange interactions it would tend to long range order with an all-in/all-out spin configuration on the tetrahedra. In fact, several theoretical and experimental studies have mostly found antiferromagnetic exchange on geometrically frustrated lattices; however, the first discovery of strong frustration in Ho$_{2}$Ti$_{2}$O$_{7}$, a ferromagnetic pyrochlore\cite{Harrisbramwell}, has challenged our understanding of cooperative magnetic phenomena. The absence of  long range magnetic ordering down to 50 mK in a Ho$_{2}$Ti$_{2}$O$_{7}$ bulk crystal \cite{HARRIS1,Harrisbramwell}, was confirmed by muon spin relaxation ($\mu$SR) and neutron-scattering experiments, but the Curie-Weiss temperature $\theta_{CW}$ was estimated to be around 2 K. \cite{HARRIS1,Harrisbramwell,Bovo, Cornelius} Furthermore, the measurement of the electronic component of magnetic heat capacity \cite{Bramhartog} has also revealed a broad magnetic peak at around 1.9 K, indicating the build-up of magnetic correlations and removal of magnetic entropy as the system is cooled. Apart from some local spin dynamics\cite{Ehlers} this leads to the spins freezing into a state analogous to ice \cite{Paul, Pauling}, a two-in/two-out configuration on the tetrahedra below T = 0.65 K \cite{krey} while retaining the nonzero Pauling entropy, $\textit{S}_{0}\approx(\textit{R}/2)\texttt{ln}(3/2)$, a key thermodynamic characteristic of frustration and the spin ice state.\cite{Bramhartog,Paul,Cornelius}

As described in \cite{hertog,Siddharthan}, it is the competition between classical nearest neighbor dipolar interactions and nearest neighbor quantum superexchange interactions that determines the degree of order (or frustration) in the material. As reported by others, a pyrochlore material is a spin ice when J$_{nn}$/D$_{nn}$ $>$ -0.91 and reverts to an unfrustrated ordered antiferromagnetic state for J$_{nn}$/D$_{nn}$ $< $-0.91 (Ho$_{2}$Ti$_{2}$O$_{7}$, J$_{nn}$/D$_{nn}$ = -0.27,  Dy$_2$Ti$_2$O$_7$ J$_{nn}$/D$_{nn}$ = -0.49) \cite{hertog, zhou, wiebe}. Reports on chemical and hydrostatic pressure studies indicate that the magnetic properties and monopole density in pyrochlore single crystals \cite{zhou, zhou2, wiebe} can be tuned via structural means. Specifically, a tetragonal distortion has been predicted to result in a symmetry breaking transition out of the Coulomb phase accompanied by a divergence of the susceptibility \cite{PhysRevLett.105.087201}, and uniaxial pressure studies performed on single crystals of Dy$_{2}$Ti$_{2}$O$_{7}$ showed that applying pressure along the [001] and [110] crystallographic directions may change the degeneracy of the ground state, inducing a 4\% increase in the magnetization at low field and a 1\% decrease in magnetization at high field for {\bf$P,H$} $\parallel$ [001] in line with predictions \cite{MITO200e4372,PhysRevLett.105.087201}. 

Epitaxial strain is usually biaxial and is fundamentally different from hydrostatic and chemical pressure. Therefore, it can be used to lower the local symmetry and to apply an effective transverse field, which should either push the system to a more ordered state or, as predicted in ref. \cite{Julia}, it will introduce quantum fluctuations in this otherwise classical system. Experimental studies on thin films of Dy$_{2}$Ti$_{2}$O$_{7}$ grown on (110) Y$_{2}$Ti$_{2}$O$_{7}$ found evidence of a zero entropy state at low temperature, suggesting that epitaxial strain can be used to tune the physics of frustrated pyrochlore magnets \cite{BovoMoya}. A study performed on Ho$_{2}$Ti$_{2}$O$_{7}$ thin films grown on yttria-stabilized-zirconia (YSZ) substrates of various orientations showed a plateau in magnetization (similar to behaviors observed in single crystals \cite{krey,Melko2004MonteModel,petrenko,Fukazawa,Cornelius}) when the field is applied along an \emph{in-plane} [111] direction, indicating that spin ice physics was preserved in those films \cite{Leusink}. It is interesting to note that from recent simulations on model spin ice thin films it is clear that not only strain, but also confinement, may impact spin ice physics in thin films. For films with surfaces perpendicular to the [001] the existence of charges at the film surface yielding an emergent square ice have been predicted \cite{PhysRevLett.118.207206}. While surface charges are predicted for films with (001) surfaces, in films with (110) and (111) surfaces confinement leads to different behaviors \cite{PhysRevX.8.021053}. In other words rich phase diagrams are expected for strained and confined versions of the spin ice.

In this paper we present structural and magnetic characterization of a thickness series of high quality epitaxial thin films of Ho$_{2}$Ti$_{2}$O$_{7}$ grown on (111), (001), and (110) YSZ substrates. We find that films grown on different orientations of YSZ lead to different strain states in the films. However it is important to note that the \emph{relaxed} portion of the films have very similar lattice parameters as extracted from x-ray diffraction (XRD) and transmission electron microscopy (TEM) measurements, regardless of the film orientation. Furthermore, this relaxed lattice parameter is on average 0.70$\%$ larger than what has been previously reported as the bulk lattice parameter (10.1 \AA \ \cite{Farmer,Ehlers2008}). Several studies have shown that the lattice parameter is quite dependent on the presence of stuffing (replacing Ti$^{+4}$ with Ho$^{+3}$, taking the system toward Ho$_{2+\delta}$Ti$_{2-\delta}$O$_{7-\delta/2}$)\cite{LAU2006, LAU2008, DTO-stuf,HTO125,Zhou_2007,Arpino}. To address the questions of stuffing, valence state, and anti-site disorder we have collected TEM images and performed XPS measurements on our thin film samples and compare these results to those collected on a single crystal. These measurements reveal the absence of stuffing and the presence of anti-site disorder and growth defects within the films. We also present magnetization measurements for all the orientations and find that in-plane and out-of-plane behaviors are different, indicating that the shape anisotropy inherent to thin films affects spin ice physics. Furthermore, while the hallmark spin ice plateau is present in a film that has the [111] direction in the plane of the film (for the (110) film), for the films with the [111] direction pointing out of the plane, the plateau is not observed. The (110) film that shows the hallmark plateau state also has less anti-site disorder compared to the other orientations. It has to be noted that the observed magnetization plateau, in the (110) film is somewhat washed out at 1.8 K. We rule out misalignment of the field with the [111] direction as a probable cause, hence, we conclude that the spin ice physics in the (110) films (and presumably in the (111) and (001) films) is modified in that they exhibit a reduction in the temperature at which correlations drive the system into the spin ice state; the observed anti-site disorder and the enlarged unit cell likely play a role.

\section{EXPERIMENTAL DETAILS}
\noindent Ho$_{2}$Ti$_{2}$O$_{7}$ epitaxial thin films were grown on (111), (001), and (110) YSZ substrates using pulsed laser deposition (PLD). The laser (KrF excimer, $\lambda = 248 \ \textnormal{nm}$) pulses were focused onto a rotating polycrystalline Ho$_{2}$Ti$_{2}$O$_{7}$ stoichiometric target with an energy density of $\sim$ 0.5 J/cm\textsuperscript{2} and repetition rate of 4 Hz. The films were grown at an average growth rate of about 0.030 nm/sec in a vacuum system with a base pressure of $1 \times 10\textsuperscript{-7}$ Torr in a 0.1 Torr oxygen background pressure while the substrate was kept at a temperature of 800~$^\circ$C. After deposition, the films were cooled to room temperature at a rate of 10~$^\circ$C/min in the same background pressure of oxygen. Atomic force microscopy using an Asylum MFP-3D system \cite{disc} operated in tapping mode was done on the substrates before growth and on the films after growth (see Supplemental Material \cite{suppmat}). After growth the surface morphology of the films show the film surfaces to be flat and smooth, with a fine grainy structure and a root mean square roughness $\sim$ 1 nm consistent with results reported by others.\cite{Leusink} Film thicknesses were determined via x-ray reflectivity measurements by fitting the thickness fringes \cite{Bjorck} (see Supplemental Material \cite{suppmat}).

\begin{figure}[t]
\centering
\includegraphics[width=3.2 in]{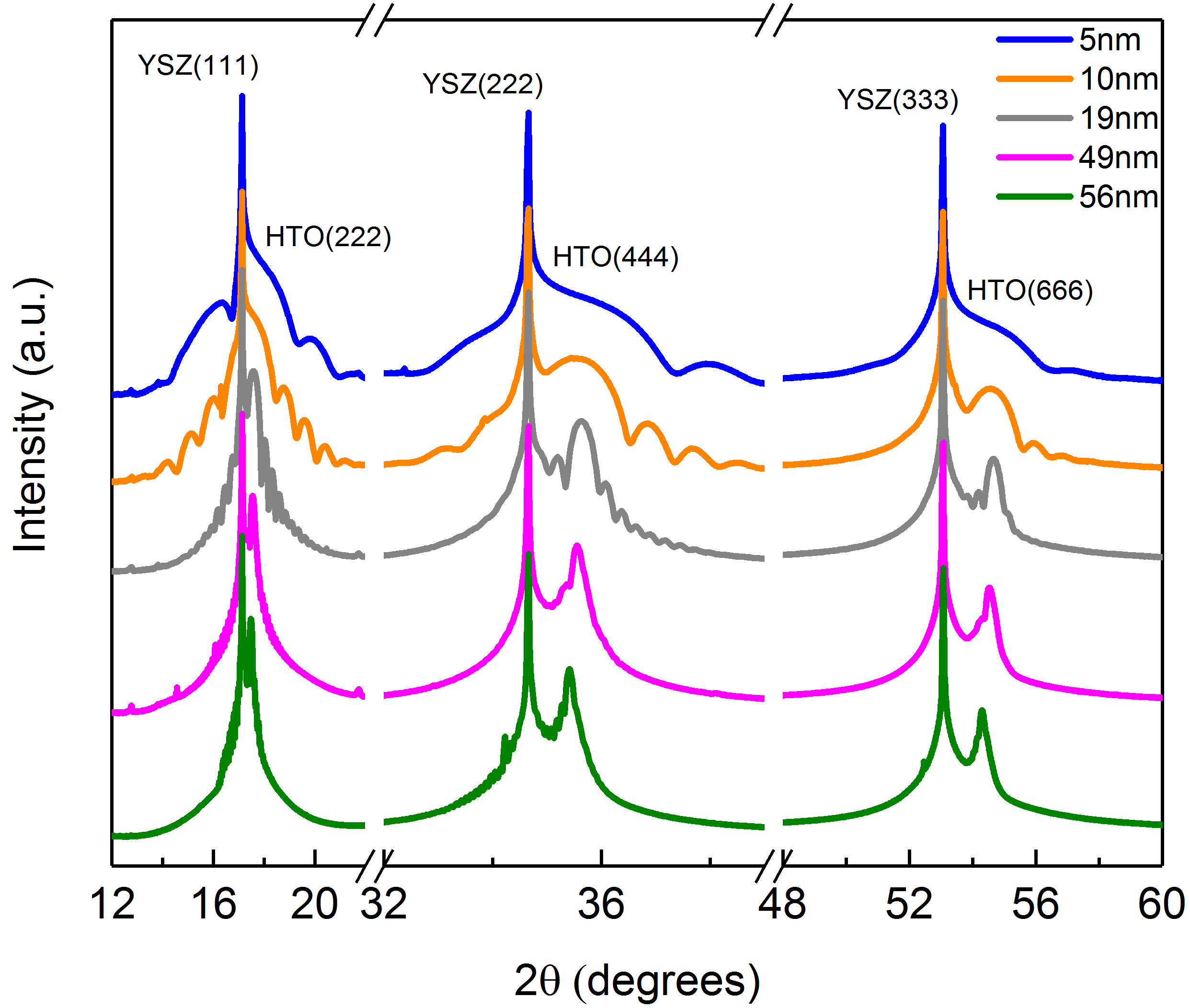}
 \caption{\label{char}(Color online) Synchrotron XRD $\theta-2\theta$-scans for (111) films ranging in thickness from 5 - 56 nm grown on YSZ substrates showing the (111), (222) and (333) substrate peaks and the (222), (444) and (666) film peaks.
}
\end{figure}

The structural characterization of the (111) films with various thicknesses was performed using synchrotron x-ray diffraction at beam line 7-2 of the Stanford Synchrotron Radiation Lightsource (SSRL) (Fig. \ref{char} a)). Reciprocal space maps (RSMs) on (001), (110), and (111) films were performed using a lab-based x-ray diffractometer at the Stanford Nano Shared Facilities at Stanford University. X-ray photoemmision spectroscopy (XPS) spectra were collected on single crystal and thin film samples using a PHI 5100 Series XPS spectrometer \cite{disc}. XPS measurements were performed using a non-monochromatic Mg $K_{\alpha}$ radiation source operated at 300 W with a 180 degree spherical sector analyzer operating at a constant pass-energy of 44.75 eV. Measurements were collected at a base pressure of $1 \times 10\textsuperscript{-10}$ Torr and at a take-off angle of 45\textdegree. Neutron diffraction experiments on the (111) films were performed at the National Institute of Standards and Technology Center for Neutron Research (NIST-NCNR) at the Multi Axis Crystal Spectrometer (MACS) \cite{macs} and the Spin Polarized Inelastic Neutron Spectrometer (SPINS) beamlines. Elastic measurements on MACS were made at a fixed neutron energy of 5 meV in double focusing mode with two cooled Be filters to eliminate higher order contamination to the beam. SPINS data was collected at a fixed neutron energy of 5 meV in flat analyzer mode using a $^3$He point detector with dual cooled Be filters at a collimation of 80'-s-80'. MACS and SPINS data reduction were performed using the DAVE software package \cite{dave}. Transmission Electron Microscopy (TEM) experiments were performed using a JEM-ARM200cF transmission electron microscope \cite{disc}. The investigated samples were prepared with a Gatan Precision Ion Polishing System (PIPS) \cite{disc}.  Magnetization measurements in an applied field of H = 1000 Oe as a function of temperature and as a function of field at fixed temperature were performed in a Quantum Design magnetic properties measurement system (MPMS) \cite{disc}.

\begin{figure}[t!]
\centering
\includegraphics[width= 2.5 in]{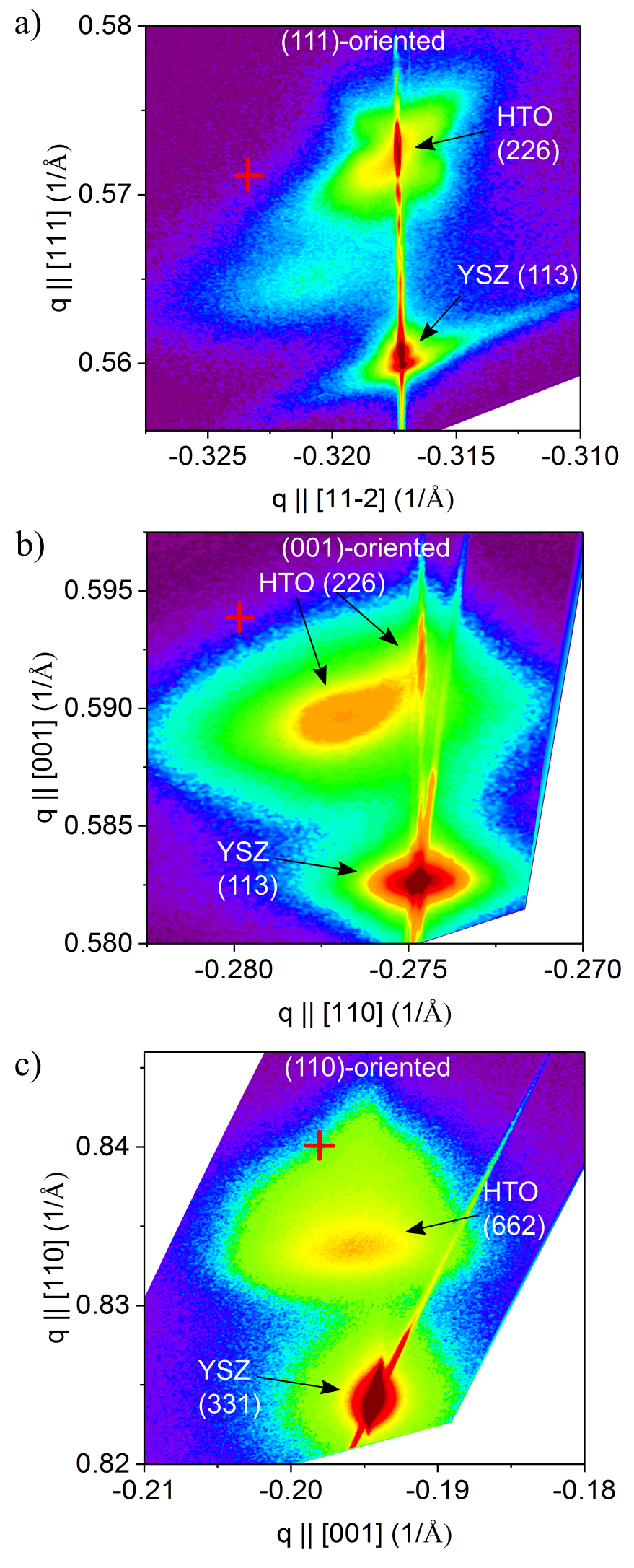}
\caption{\label{RSM}(Color online) a) Reciprocal space map of the (226) diffraction peak of Ho\textsubscript{2}Ti\textsubscript{2}O\textsubscript{7} for a 56 nm thick film grown on a (111) YSZ substrate. b) RSM of the (226) diffraction peak of Ho\textsubscript{2}Ti\textsubscript{2}O\textsubscript{7} for a 50 nm thick film grown on a (001) YSZ substrate. The (113) diffraction peak of YSZ is marked by the intensity at the bottom of each image. c) RSM of the (662) diffraction peak of Ho\textsubscript{2}Ti\textsubscript{2}O\textsubscript{7} for a 118 nm thick film grown on a (110) YSZ substrate. The (331) diffraction peak of YSZ is marked by the intensity at the bottom of the image. The red cross hairs indicate the position of bulk Ho\textsubscript{2}Ti\textsubscript{2}O\textsubscript{7} with a lattice parameter of 10.1 \AA \ in each image. The diagonal streaks of intensity across the substrate reflection in b) and c) are a result of the finite step size of the one-dimensional detector used for the measurement.}
\end{figure}

\begin{figure*}[t]
\centering
\includegraphics[width =0.8 \textwidth]{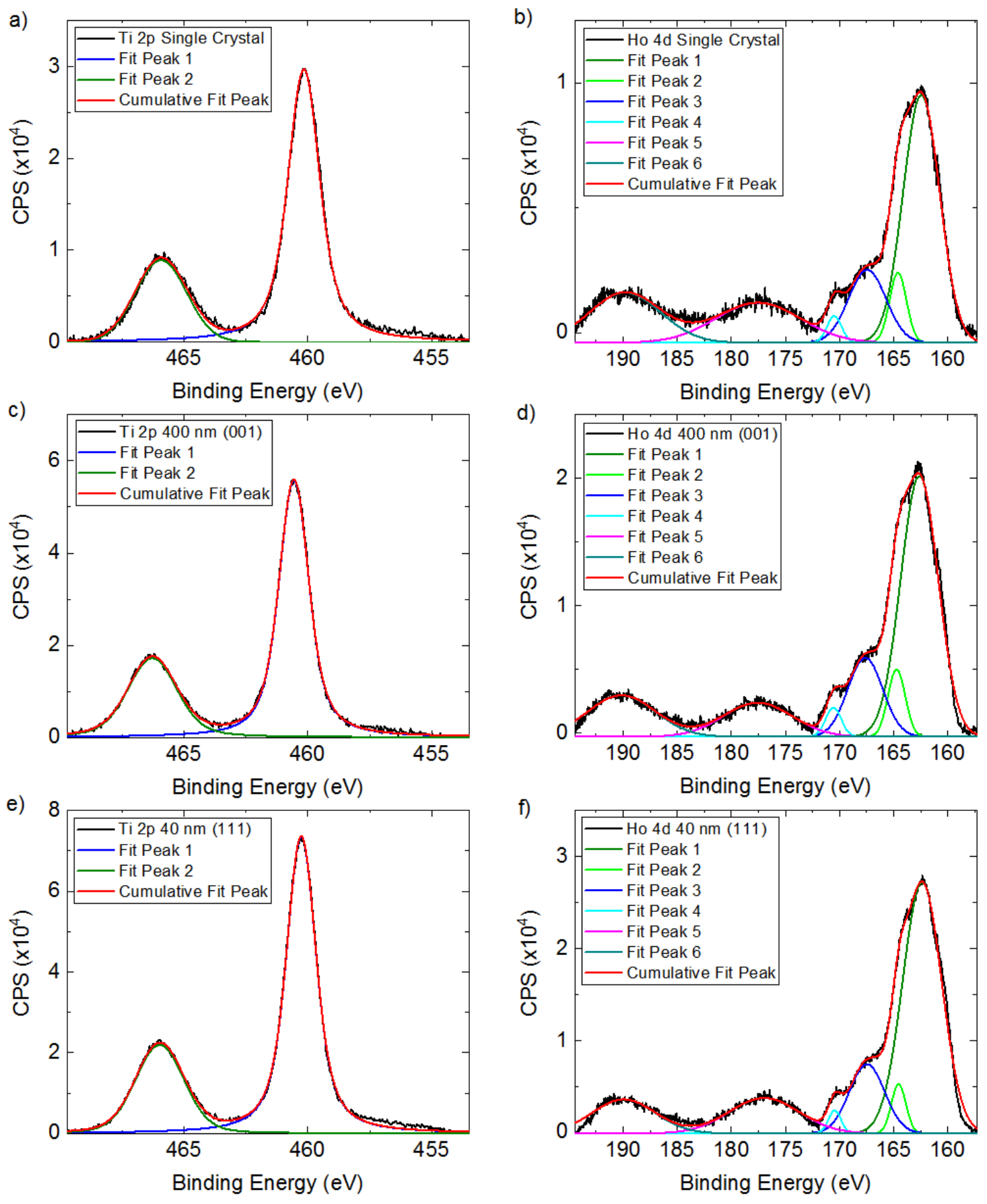}
\caption{\label{XPS}(Color online) XPS spectra of Ti 2p peaks and Ho 4d peaks for a Ho\textsubscript{2}Ti\textsubscript{2}O\textsubscript{7} single crystal, a 400 nm (001) thin film, and a 40 nm (111) thin film. a) Ti 2p region of single crystal. b) Ho 4d region of single crystal. c) Ti 2p region of 400 nm (001) thin film. d) Ho 4d region of 400 nm (001) thin film. e) Ti 2p region of 40 nm (111) thin film. f) Ho 4d region of 40 nm (111) thin film. The best fits for each set of spectra are displayed along with the baseline subtracted data from the actual scan. For more details on baseline subtraction see the Supplemental Materials \cite{suppmat}}
\end{figure*}

\section{RESULTS AND DISCUSSION}

\noindent The results of the structural characterization done at SSRL are presented in Fig.~\ref{char}, which shows $\theta-2\theta$ scans near the YSZ (111), (222), and (333) reflections for Ho$_{2}$Ti$_{2}$O$_{7}$ thin films grown on (111) YSZ of varying thickness between 5 and 56 nm. Prominent thickness fringes can be observed on all the major film peaks, displaying the high structural quality and thickness uniformity of our films. A trend of decreasing $2\theta$ position with increasing film thickness can be observed for all film reflections, indicating an increase of the out-of-plane (111) d-spacing with increasing film thickness. 

To investigate the strain state of our films, RSMs were measured on films grown on (111), (001), and (110) YSZ substrates. The RSMs taken around the (226) diffraction peak of Ho$_{2}$Ti$_{2}$O$_{7}$ for a 56 nm (111) and a 50 nm (001) thin film, and around the (662) for a 118 nm (110) thin film are presented in Fig. \ref{RSM} a) - c). The previously reported lattice parameter of bulk Ho$_{2}$Ti$_{2}$O$_{7}$ is 10.1 \AA\cite{Farmer,Ehlers2008} while the lattice parameter of YSZ is 5.15  \AA \ as extracted from our synchrotron measurements presented in Fig. \ref{char} and reports by others\cite{Yashima,Pomfret}, yielding a lattice mismatch of about 2\% between film and substrate (tensile strain) with one Ho$_{2}$Ti$_{2}$O$_{7}$ unit cell fitting onto a block of 2x2 YSZ unit cells. We find that for the 56 nm film grown on (111) YSZ most of the scattering intensity from the film occurs at an in-plane value of the scattering vector matching that of the substrate; in other words, most of this film has in-plane lattice parameters that are matched with the YSZ, i.e., this portion of the film is fully strained. The film does show some signs of strain relaxation, i.e., that the top layer of the film has relaxed (intensity appears at different in-plane q-values). The 50 nm (001) film displays a strained interfacial layer, but it is clear that a significant portion of the film exhibits in-plane values for the scattering vector that are distinct from that of the substrate, indicating the presence of considerable relaxation. Here we note that in the previous work done \cite{Leusink}, it was found that no intermediate strained layer was observed for a 33 nm (001) film studied. In Fig. \ref{RSM} c) an RSM around the (662) Ho$_{2}$Ti$_{2}$O$_{7}$ reflection of a 118 nm film grown onto (110) YSZ is shown, this RSM does not show evidence of a strained interfacial layer, but only a single presumably fully relaxed film (see Supplemental Material\cite{suppmat} for additional RSMs for a 132 nm (111) film, and for a 35 nm (001) film). In each RSM we have also shown the position of where the bulk Ho$_{2}$Ti$_{2}$O$_{7}$ reflection would occur based off of the previously reported lattice parameter of 10.1 \AA. It is clear that our films are not relaxing to this previously reported bulk value. Using the q-values from films grown on each substrate orientation we extract lattice parameter values for both the strained and relaxed portions of the films (see Table \ref{table:1} and the Supplemental Material for more details \cite{suppmat}). The lattice parameters extracted from this analysis are similar to those previously reported for stuffed variants of HTO, when 30$\%$ of the Ti$^{4+}$ sites have been replaced with Ho$^{3+}$\cite{LAU2006, LAU2008,HTO125,Zhou_2007,Arpino}. However, these stuffed stoichiometries are reportedly difficult to obtain and have been synthesized previously via a rapid cooling method from temperatures much higher than that employed in our growth routine, making them unlikely to be thermodynamically accessible.

Selected area electron diffraction (SAED) patterns were collected using a TEM for films grown on each substrate orientation. We find bright and well resolved diffraction spots in each case. The SAED pattern for a 70 nm (111) film is shown in the Supplemental Material as a typical example \cite{suppmat}. This SAED pattern was used to extract both the $a$ and $c$-axis lattice parameters for the (111) film away from the substrate interface. The extraction process is presented in the Supplemental Material \cite{suppmat} and the resulting values are presented in Table \ref{table:1} along with lattice parameters extracted from x-ray diffraction. The lattice parameters from SAED are also consistent with an inflated unit cell compared to what has been previously reported for bulk single crystals. For the films grown on (001) YSZ we can explicitly determine that the unit cell is tetragonal. The films grown on (111) YSZ are rhombohedral while those grown on (110) YSZ are orthorhombic. Using the values presented in Table \ref{table:1} it was found that the rhombohedral angle deviates from 90$\degree$ by $\sim$+$0.2\degree$ in the (111) case, consistent with the in-plane tensile strain, while in the (110) film the angle between the $a$($b$) axis and the [110] direction was found to be $\sim$ 45.1$\degree$.
%\vskip0.1in
\begin{table*}[t]
\caption{Lattice Constant Summary. For the (001) film the unit cell is tetragonal with the $c$-axis normal to the film plane, while the $a$-axis is in the film plane. For the (110) films the $c$-axis is an in-plane direction while the [110] direction points out of the film plane. The lattice parameters for these films were determined from x-ray diffraction (assuming a tetragonal unit cell for the (110) film). For the (111) film the $c$-axis and $a$-axis parameters are determined from ($00l$) and ($hh0$) reflections from the TEM SAED patterns (assuming a tetragonal unit cell) taken some distance removed from the substrate interface; these parameters are likely associated with a partially or fully relaxed region of the film . 
The error bars were determined based on the FWHM values of the reflections (see Supplemental Material for more details \cite{suppmat}). } 
\centering
%\resizebox{0.5\textwidth}{!}{%
\begin{tabular}{|c |c | c | c | c | c |} 
 \hline
Film & Film & $c$ &$c$  &$a$ & $a$ \\
Thickness & Normal & (strained)&  (relaxed)& (strained) & (relaxed) \\
 \hline
50 nm & [001]& 10.13(3) & 10.17(2) &  10.29(3) & 10.21(6) \\\hline
118 nm & [110]& - & 10.2(2) &- & 10.18(1)  \\\hline

    70 nm & [111]& \multicolumn{2}{c|}{10.15(2)}&   \multicolumn{2}{c|}{10.13(6)} \\\hline

\end{tabular}
\label{table:1}
\end{table*}

\begin{figure*}
\centering
\includegraphics[width=1.0\textwidth]{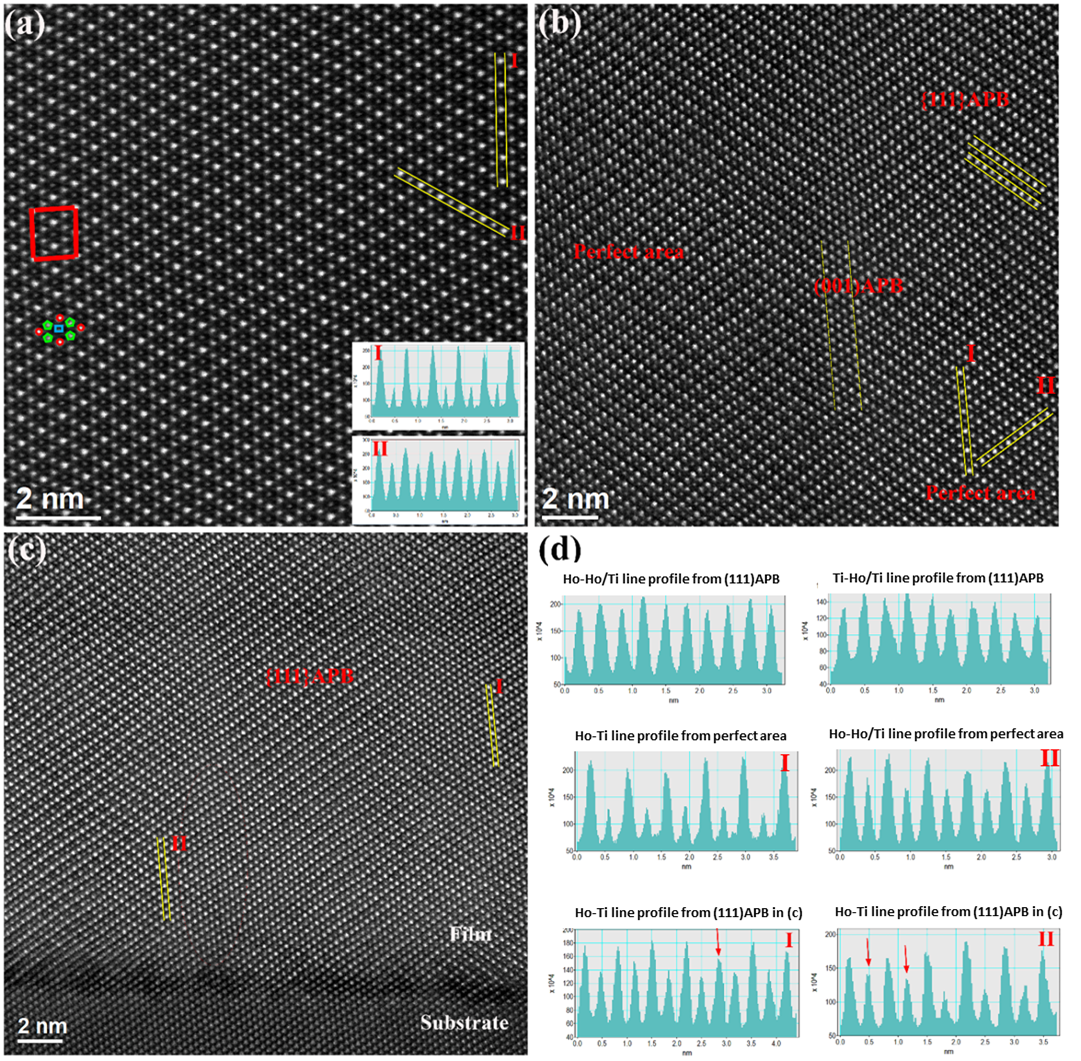}
\caption{\label{TEM2}(a) HAADF-STEM image of Ho$_2$Ti$_2$O$_7$ single crystal along [110]. The unit cell is delineated with a red rectangle. Pure Ho atomic column indicated by red circle; Ho/Ti mixed column indicated by green polygon; Pure Ti column indicated by blue rectangle.  Insets: Intensity line profiles of Ho-Ti (I) and Ho-Ho/Ti (II). 
(b) HAADF-STEM image of the 118 nm (110) Ho$_2$Ti$_2$O$_7$ film collected away from the film/substrate interface.
(c)	HAADF-STEM image of the 118 nm (110) film at the film/substrate interface region. 
(d)	Intensity line profiles from atomic columns indicated by yellow lines in (b) and (c): Top: Ho-Ho/Ti and Ti-Ho/Ti from {111}APB of (b); middle: Ho-Ti and Ho-Ho/Ti line profiles from perfect area in film of (b); Bottom: Ho-Ti line profiles from (c) showing anti-sites disorder. Red arrow indicates Ti into Ho sites in (I) and Ho into Ti sites in (II).
}
\end{figure*}

In order to investigate whether these differences in lattice parameter compared to the previously reported bulk value could be due to deviating stoichiometry, or oxygen deficiency, we have performed XPS measurements to confirm a 1:1 Ho:Ti ratio in our films. The resulting XPS spectra of the Ti 2p and Ho 4d regions of a Ho$_{2}$Ti$_{2}$O$_{7}$ single crystal, a 400 nm (001) thin film, and a 40 nm (111) thin film are presented in Fig. \ref{XPS}. The Ti $2\mathrm{p}_{\frac{1}{2}}$ and Ti $2\mathrm{p}_{\frac{3}{2}}$ spin-orbital split doublet was clearly observed for both the thin film and single crystal samples, and binding energies are similar to those previously reported for $\mathrm{TiO}_{2}$ \cite{Simson,McCafferty,Bedri_E,Bukauskas_V}. The spectral contributions from the Ti $2\mathrm{p}_{\frac{1}{2}}$ and Ti $2\mathrm{p}_{\frac{3}{2}}$ doublet were quantified utilizing a mixture of Lorentzian and Gaussian curves for fitting. The multiplet split spectra of Ho 4d has been observed for the single crystal and the thin film samples. In each case the spectra, while yielding no well defined peaks, is qualitatively similar to previous measurements reported on holmium oxide with features occurring at similar binding energies to those reported in the range from 200 eV to 160 eV \cite{Ogasawara,Lang,Teterin,Padalia}. The different spectral contributions from the Ho 4d multiplet structure were determined by fitting six distinct Gaussian peaks to each data set. Observed differences in the binding energies between the single crystal and the thin films for the two regions, though small, can be attributed to different amounts of sample charging during each measurement. The extracted areas from the fitting of the Ti 2p and Ho 4d peaks as well as their respective sensitivity factors were used to calculate the relative concentrations of Ho and Ti in each sample. The concentrations were found to be Ti: $50.25 \ \pm{0.35} \%$, Ho: $49.75 \ \pm{0.48} \%$ for the single crystal, Ti: $49.75 \ \pm{0.43} \%$, Ho: $50.25 \ \pm{0.48} \%$ for the 400 nm (001) thin film, and Ti: $49.67 \ \pm{0.33} \%$, Ho: $50.33 \ \pm{0.41} \%$ for the 40 nm (111) thin film. Here the error bars on each concentration are derived from the fitting error on each extracted area. Hence, the films look identical to the single crystal, i.e., they show 1:1 Ho:Ti stoichiometry, and the Ti 2p peaks are consistent with Ti$^{4+}$ for all measured samples ruling out mixed valence on the Ti-site. A detailed example of the curve fittings and concentration calculations is presented in the Supplemental Material \cite{suppmat}. 

To further investigate the microstructure of the films we compare high-angle-annular-dark-field scanning TEM (HAADF-STEM) images of a HTO single crystal and of films grown on each substrate orientation ((111), (001), and (110)). Fig. \ref{TEM2} (a) shows an atomic resolution image of the single crystal, while Fig. \ref{TEM2} (b) and (c) show images from the 118 nm (110) HTO film, collected away from, and close to, the substrate interface, respectively. All images are taken viewed along $<$110$>$ because along this direction columns of pure Ho and pure Ti, as well as mixed Ho/Ti columns can be observed. The characteristic atomic column intensity profiles along two different directions, across Ho-Ti columns (I) and across Ho-Ho/Ti columns (II) for the single crystal sample are shown in the inset in Fig. \ref{TEM2} (a), i.e., the pure Ho (Z = 67) columns have a much larger intensity compared to the pure Ti (Z = 22) columns.

From the images taken on the (110) film we can easily identify that the film has anti-phase boundaries (APBs), typical of those found in pyrochlore structures \cite{xin},these are very likely growth defects. In Fig. \ref{TEM2} (d) we show the atomic column intensity profiles for the (110) film for the two different regions that were studied. Away from the substrate interface (Fig. \ref{TEM2} (b)) we present the profiles for a (111) APB (top panels in Fig. \ref{TEM2} (d), across Ho-Ho/Ti and Ti-Ho/Ti columns), and for a "perfect area" (center panel in Fig. \ref{TEM2} (d), across Ho-Ti columns and across Ho-Ho/Ti columns). The profiles of the "perfect area" show intensities very comparable to the single crystal, clearly showing that a large region of the (110) film shows a crystallinity and stoichiometry expected for pristine Ho$_2$Ti$_2$O$_7$. Fig. \ref{TEM2} (c) shows a region close to the YSZ substrate. Two profiles (I and II, both taken across Ho-Ti columns) are presented in the bottom panel of Fig. \ref{TEM2} (d). There is little intensity difference between Ho and Ti columns along one [110]  direction, unlike the alternating intensity of the perfect area (center panel). This indicates that in addition to the APBs, there is clear presence of anti-site disorder, as illustrated by the intensity variations of pure Ho and Ti columns (see red arrows) in Fig. \ref{TEM2} (d) bottom panel. In the intensity line profile I (across Ho-Ti columns), one Ho column has lower intensity than the adjacent Ho columns indicating quite a few Ti went into this column, and vice versa, the two Ti columns have higher intensity, implying Ho went into the Ti columns in line profile II. We have repeated a similar study regarding APBs and anti-site disorder on a (111) and on a (001) film, which we present in the Supplemental Material \cite{suppmat}. A comparison of the three films to the single crystal clearly indicates that the films have APBs and anti-site disorder, but there is no evidence of an appreciable amount of stuffing, i.e., additional Ho on the Ti-site. 

\begin{figure*}
\centering
\includegraphics[width=0.85\textwidth]{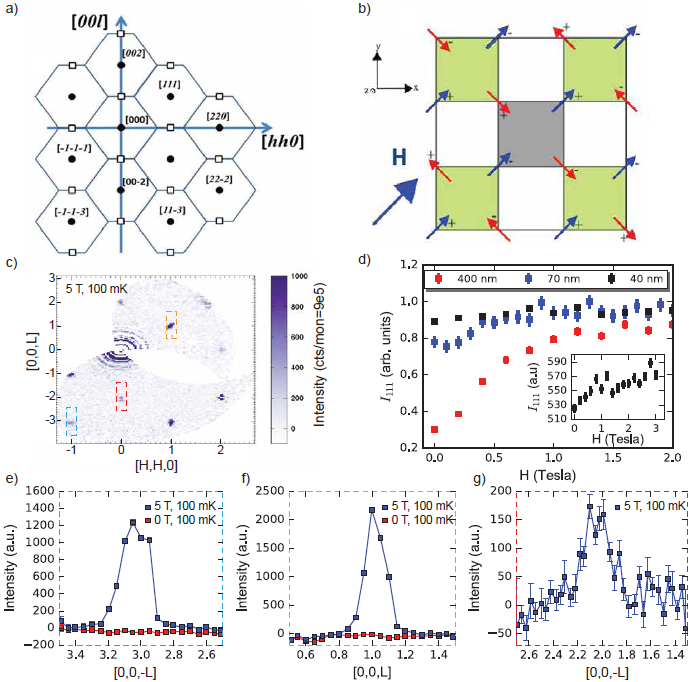}
\caption{\label{Neutron}(Color online) a) A representation of the reciprocal space map for the pyrochlore lattice. The full circles show the positions of Bragg peaks for a $Q= 0$ (zone center) magnetic structure, while the open squares show the $Q = X$ positions (zone boundary) \cite{Harrisbramwell}. b) The Ho$_2$Ti$_2$O$_7$ unit cell projected down the z-axis. The component of each spin that is parallel to the z axis is designated by the + and - signs. The four tetrahedra that make up the unit cell appear as the light green squares. The grey square in the middle is not a tetrahedron, but its diagonally opposite spins inhabit the same lattice plane. The schematic depicts the $Q = X$ magnetic structure (antiparallel $\beta$-chains (red) and polarized $\alpha$-chains (blue)). The large arrow along the [110] direction indicates the direction of the applied magnetic field. c) Map of the scattering intensity of the ($hhl$) plane of a 400 nm thick Ho$_{2}$Ti$_{2}$O$_{7}$ sample grown on (111) YSZ after field cooling the sample to T = 0.1 K in H = 5 T applied along the [1-10] direction (in-plane) of the film. To obtain this map the scattering intensity at T = 25 K and H = 0 T was subtracted as background. d) Integrated intensity of the [111] magnetic film peak as a function of applied field at T = 100 mK for a 400 nm, 70 nm, and 40 nm thick (111) film normalized to the intensity at 2 T. (inset) Unnormalized data from the 40 nm film to illustrate small intensity increase with field. For these neutron measurements, all error bars and confidence intervals are given by standard deviations of the Poisson distribution. Cuts taken along the (00$l$) direction through the e) ($\bar{1}$$\bar{1}$$\bar{3}$) and f) (111) reciprocal space positions integrated along ($hh$0) from (0.9 0.9 $l$) to (1.1 1.1 $l$) at 100 mK and 5 T (blue) for the data in (c) and for an equivalent data set taken at 100 mK and 0 T (red). The zero field data set has the same 25 K and 0 T data set subtracted as background. g) Cut taken along the (00$l$) direction through the (00$\bar{2}$) reciprocal space position integrated along ($hh$0) from (-0.15 -0.15 $l$) to (0.15 0.15 $l$) for the background subtracted data set in (c).}
\end{figure*}
A series of neutron scattering measurements have been performed on (111) Ho$_2$Ti$_2$O$_7$ thin films with thicknesses ranging from 40-400 nm using the MACS and SPINS instruments at the NCNR. Schematic representations of the scattering map and the corresponding magnetic structure of the Ho$_2$Ti$_2$O$_7$ unit cell are depicted in Figs. \ref{Neutron} a) and b). Fig. \ref{Neutron} c) shows a scattering map of the scattering intensity of the ($hhl$) plane of the 400 nm film taken at T = 0.1 K on MACS after field cooling the film in a 5 T field (applied along the [1$\bar{1}$0] direction), with data at 25 K and 0 field subtracted to more clearly show the magnetic contribution. With the field applied along the [1$\bar{1}$0] direction of Ho$_2$Ti$_2$O$_7$, as described in \cite{Harrisbramwell}, the spin system is separated into two sets of chains, parallel to the field ($\alpha$-chains), and perpendicular to the field ($\beta$- chains) (see Figs. \ref{CS12} a) and \ref{Neutron} b)). The $\beta$ chains can either be parallel or anti-parallel to one another. The parallel case gives scattering intensity at $Q = 0$ while the anti-parallel case, associated with a symmetry lowering, will result in scattering intensity at the $Q = 0$ and the $Q = X$ positions (see Fig. \ref{Neutron} a) and b)), with zone boundary peaks reportedly increasing in intensity in applied fields \cite{clancy}.

In the scattering map presented in Fig. \ref{Neutron} c), the intensity near the (22$\bar{2}$) and (222) positions is due to the (11$\bar{1}$) and (111) YSZ structural reflections respectively. The intensities at the ($\bar{1}$$\bar{1}$$\bar{3}$), (11$\bar{3}$), (111), and ($\bar{1}$$\bar{1}$$\bar{1}$) positions are from the film and are dramatically enhanced when a magnetic field is applied, hence they are attributed to the magnetic order in the bulk of the film associated with the polarized state. Fig. \ref{Neutron} d) shows the scattering intensity of the (111) film reflection as a function of applied field for films of varying thickness taken on SPINS. The data is normalized by the intensity measured at 2 T. The results indicate that the films saturate into the high-polarized state under applied fields and that lower saturation fields are needed for the thinner films. The intensity at the (002) and (00$\bar{2}$) film positions only shows up under applied field. Figs. \ref{Neutron} e) and f) show cuts along the (00$l$) direction through the ($\bar{1}$$\bar{1}$$\bar{3}$) and (111) reciprocal space positions respectively integrated from (0.9 0.9 $l$) to (1.1 1.1 $l$) at 100 mK at 0 T (red) and at 5 T (blue), with the same 25 K and 0 T scans subtracted as background. Fig. \ref{Neutron} g) shows the same cut through the subtracted (00$\bar{2}$) at 100 mK and 5 T. These illustrate the dramatic magnetic enhancement at these reflections, as we see no difference for the film nuclear peaks from 25 K to 0.1 K in zero field (i.e. no peak), but see clear peaks when the field is applied. The (002)-like reflections in particular are not visible at all in the unmagnetized state, indicating that these peaks are magnetic origin.

To summarize, the presence of only the (113), (111), and (002)-like film peaks, i.e., $Q = 0$, after field cooling the film in a 5 T field, and the absence of scattering at the $Q = X$ positions, indicates that the 400 nm film only shows parallel $\beta$-chains and polarized $\alpha$-chains (see Fig. \ref{Neutron} b)). However, the absence of scattering at these positions could be a result of substrate background issues that hinder the observation of diffuse scattering. Important to note is that the appearance of the (002) is consistent with the Q = 0 magnetic structure as previously reported \cite{Harrisbramwell,Fennell, clancy}.

\begin{figure}
\centering
\includegraphics[width= 3.1 in,keepaspectratio]{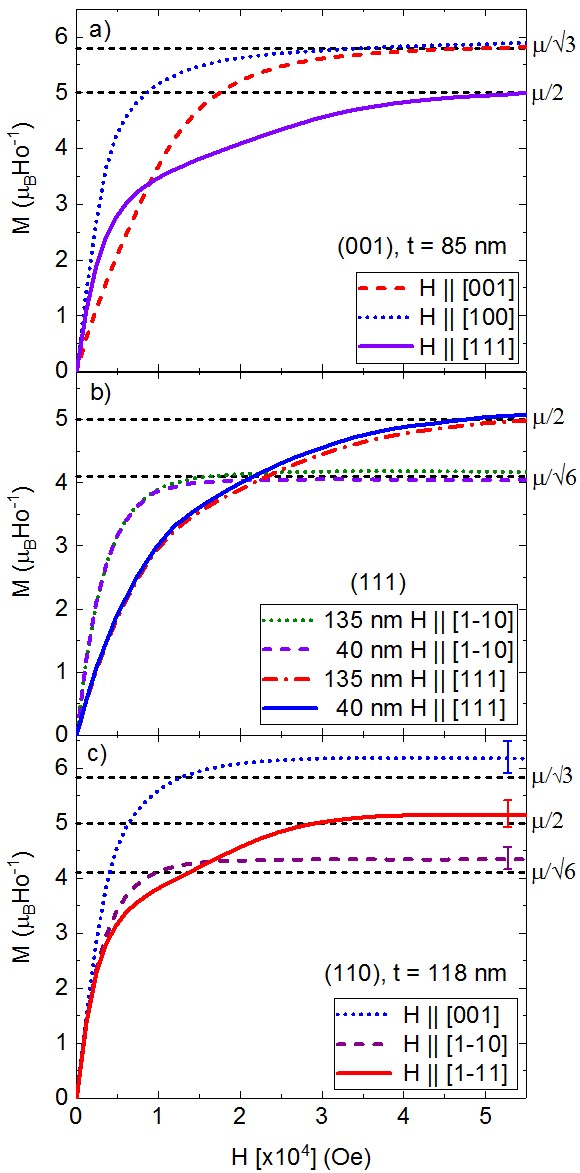}
\caption{\label{MvsH}(Color online)  Magnetization plotted as a function of applied field for a) 85 nm thick (001) thin film with H applied along the [001] (out-of-plane), [100] (in-plane), and [111] (54.7$^{\circ}$ from the film normal) directions. b) For a 135 nm, and a 40 nm thick (111) thin film with H applied along the [111] (out-of-plane) and [1$\bar{1}$0] (in-plane) directions. c) For a 118 nm thick (110) thin film with the field applied along the [001], [1$\bar{1}$0], and [1$\bar{1}$1] in-plane directions. The error bars ($\pm{5\%}$) on each curve are a result of the error in the fitted thickness from the x-ray reflectivity results (see Supplemental Materials \cite{suppmat}) and are representative of the error bars for the (111) and (001) samples as well. The expected values for the saturation magnetization from simulations of the dipolar spin ice model are displayed as the black dashed horizontal lines in each figure ($\mu= 10 \ \mu_B/\mathrm{Ho}$, $\mu_B$/Ho = 927.4 x 10$^{-26}$ J T$^{-1}$/Ho). All measurements were performed at T = 1.8 K. The curves are corrected for their individual substrate contributions.}
\end{figure}

Next, we investigate the anisotropy in each of the films by discussing the magnetization as a function of applied field measured at 1.8 K (see Fig. \ref{MvsH}). It is important to emphasize that all MPMS measurements were performed with the field along the indicated crystallographic directions with a precision of about 2$^{\circ}$ (see Supplemental Material for further details \cite{suppmat}). In panel a) the magnetization is shown for a 85 nm thick film grown on (001) YSZ with the field applied along the [100] in-plane direction (dotted curve), the [001] direction (along the film normal, dashed curve) and along the [111] direction (54.7$^{\circ}$ from the film normal). The horizontal dashed lines indicate the expected saturation magnetization values obtained from simulations of the dipolar spin ice model when the field is applied along each of the different crystallographic directions of the sample ($\mu= 10 \ \mu_B/\mathrm{Ho}$)\cite{Melko2004MonteModel}. We find that for H $|| <100>$ and H $||$ [111] the film shows the expected M = 5.8 and 5 $\mu_B/\mathrm{Ho}$, respectively. Comparing the responses when H $||$ [001] and H $||$ [100] it is clear that the film saturates at significantly lower fields when the field is applied in the film plane. This shows that the geometric anisotropy intrinsic to the film may play a role in the modification of spin ice physics. The observed slow saturation of the film when H $||$ [111], which has a significant component along the film normal, may be a result of the same shape effect and likely plays a role in obscuring the magnetization plateau. The effect of shape anisotropy on the magnetic properties of HTO thin films is also evident from temperature dependent susceptibility  measurements. Below 10 K, the temperature dependent magnetization shows a clear linear regime observed only for in-plane directions, very similar to observations by others in single crystals \cite{Siddharthan}, while a more parabolic shape is observed for out-of-plane directions (see Supplemental Material for more details \cite{suppmat}). 

In panel b) the magnetization is shown for a 40 nm and a 135 nm thin film grown on a (111) YSZ substrate. For each film, data is shown for the field applied along the [111] out-of-plane direction and for the [1$\bar{1}$0] in-plane direction. We find that the saturation magnetization values correspond to the expected values for bulk (for H $||$ $[1\bar{1}$0], M = 4.1 $\mu_B$/Ho). For this orientation we again find that the films have a hard time saturating when the field is applied in the out-of-plane direction.  For these (111) films there is no evidence of a plateau state. While shape anisotropy may play a role in how easily the films are saturated, interestingly, no thickness dependence is observed, i.e., the 40 nm film, in which the strained layer makes up a significant portion of the film, and the 135 nm film, which is mostly relaxed, behave in the same way. This lack of thickness dependence and the lack of a plateau state has been observed for both (001) and (111) films (see Supplemental Material for more details \cite{suppmat}).

\begin{figure}[t]
\centering
\includegraphics[width= 3 in,keepaspectratio]{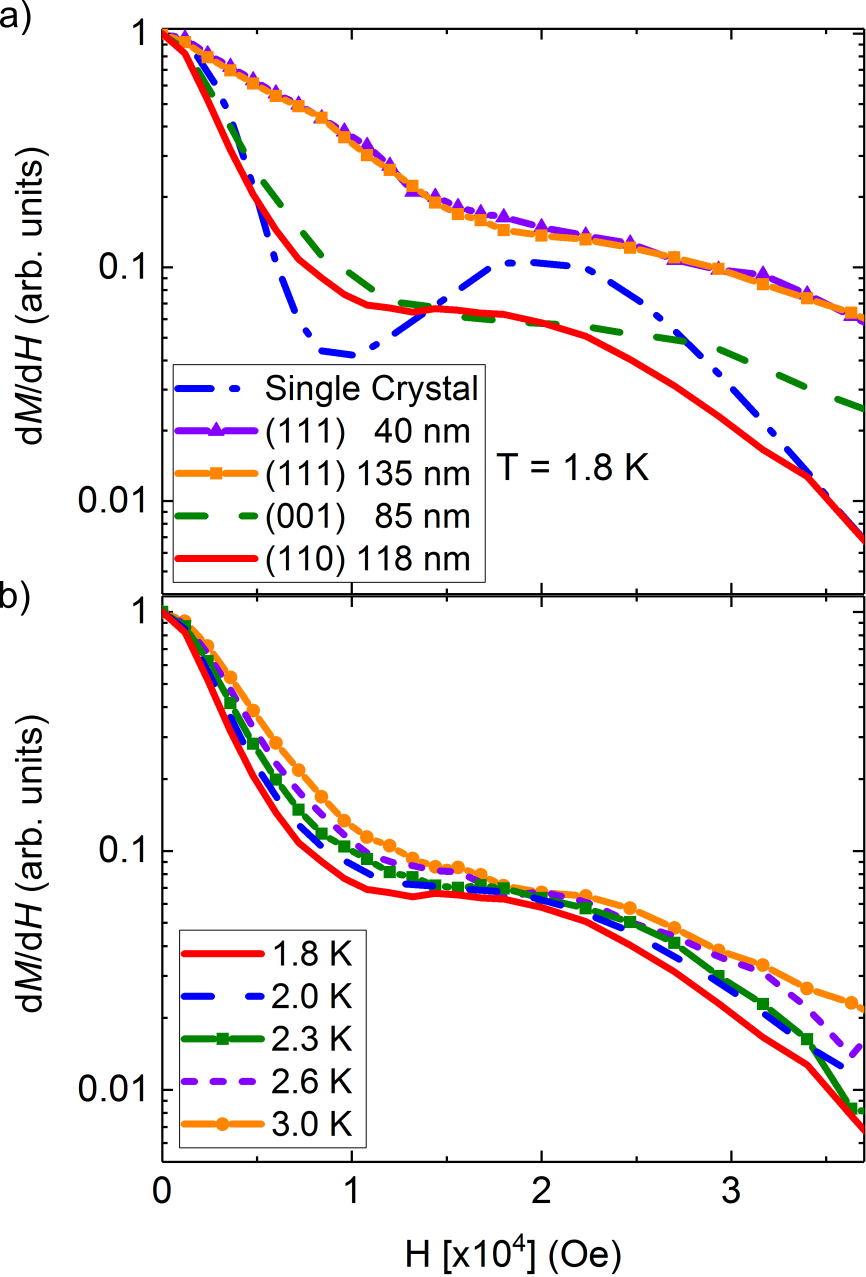}
\caption{\label{MvsHder}(Color online) a) Derivative $dM/dH$ as a function of applied field along the [111] direction for films grown on (001) YSZ, (110) YSZ, and (111) YSZ. The measurement of a single crystal with H $||$ [111] is displayed for comparison. All measurements were performed at T = 1.8 K. b) Derivative $dM/dH$ as a function of temperature for the (110) film with H $||$ [111]. The vertical axis in each figure has been normalized and is plotted on a log-scale to highlight the presence of a plateau in M vs. H.}
\end{figure}

In panel c) we show the response for a 118 nm (110) film, which has the [111] direction in the plane of the film. Here measurements with H $||$ [001], [1$\bar{1}$0], and [1$\bar{1}$1] are all in plane directions and saturate rapidly with increasing field in a way very similar to bulk observations \cite{krey,Melko2004MonteModel}. The error bars on each curve are a result of the error in the fitted thickness from the x-ray reflectivity measurements collected for the sample. This error encompasses a $\pm{5\%}$ range around the fitted values and is representative of the error expected for the other samples grown on (111) and (001) substrates as well. Although, the plateau state is quite weak at this temperature (1.8 K) it is clearly visible for the [111] direction in this film.

To further highlight the presence of the plateau state in the (110) film we present the derivatives ($\frac{dM}{dH}$) of all four films with H $||$ [111] in Fig. \ref{MvsHder} a). For comparison we also show the derivative of a measurement performed on a single crystal. The development of an intermediate plateau at around 3.33 $\mu_B$/Ho, with the onset of a magnetic transition and saturation to around 5.0 $\mu_B$/Ho, is clearly visible in our single crystal, consistent with reports by others \cite{Cornelius,petrenko,Fukazawa,krey}, and with expectations from Monte Carlo simulations of the dipolar spin ice model \cite{Melko2004MonteModel}. For the single crystal we have investigated the effect of misalignment of the field with the [111] direction. Magnetization vs. field curves for a single crystal, in which we purposely misaligned the field away from the [111] direction by 4$^{\circ}$ (both toward the [110] and [001] directions) and by 10$^{\circ}$ (towards the [001]), shows that misalignment of well over 4$^{\circ}$ is needed to significantly diminish the plateau. Note, the crystal was measured using the same system and alignment procedure as is used for the films. Furthermore, we find that the plateau in the single crystal diminishes quickly as the temperature is increased from 1.8 - 3 K (see Supplemental Material for more details \cite{suppmat} on the temperature dependence and the misalignment study). To confirm the disappearance of the plateau with increased temperature, we show a temperature dependent study on the (110) thin film plateau state in Fig. \ref{MvsHder}  b), which clearly shows that the plateau disappears when the temperature is increased by about 1 K. Based on these results misalignment in the measurement can effectively be ruled out as a cause for the smudged plateau in the (110) thin film. Therefore, we could conclude that the the spin ice physics is modified in the films leading to slight suppression of the temperature at which the spin ice state gets locked in. 
\section{CONCLUSIONS}
We have grown high quality epitaxial thin films of the pyrochlore titanate Ho$_2$Ti$_2$O$_7$ and find that the orientation of the substrate plays a key role in determining the strain state and the relaxation rate of the films. Elastic neutron scattering experiments on (111) films clearly showed the presence of the $Q=0$ phase in films of varying thickness, but observation of the $Q=X$ phase (the plateau state in magnetization measurements) in the films remains elusive. XRD and TEM measurements show that our films have an enlarged unit cell compared to the expected 10.1 \AA \, however, this is not caused by stuffing. While TEM measurements do not fully rule out the possibility of a minor amount of stuffing, there is a clear presence of anti-site disorder and APBs in the films. The increased lattice parameters is likely due to strain and the disorder in the films. Magnetization measurements further support that we have stoichiometric HTO. In our extensive magnetization studies we find a diminished, but clearly observable plateau for the (110) film (i.e., when the [111] direction is an in-plane direction), and an absence of the plateau states in the (001) and (111) films.  It is unlikely that misalignment is responsible for the absence of the plateau states in these orientations. We find that for the single crystal a misalignment of well over 4$^{\circ}$ is necessary to start losing the plateau. Looking at the similarities between the single crystal M vs. H measurement at 3 K and our thin film measurement at 1.8 K for the (110) film, it seems plausible that the diminishing of the plateau is due to a suppression of the temperature at which correlations drive the system into the spin ice state. This suppression is likely due to the presence of defects in the films, and perhaps the slightly larger separation of the Ho-ions in the films due to the inflated unit cell.

\section{ACKNOWLEDGEMENTS}
C.B. acknowledges support from the National Research Foundation, under grant NSF DMR-1847887. J.N., C.C., and T.S. acknowledge support from the National Research Foundation, under grant NSF DMR-1606952. A portion of this work was performed at the National High Magnetic Field Laboratory, which is supported by National Science Foundation Cooperative Agreement No. DMR-1157490, No. DMR-1644779, and the State of Florida. H.D.Z acknowledges support from the NHMFL Visiting Scientist Program, which is supported by NSF Cooperative Agreement No. DMR-1157490 and the State of Florida. Part of this work was performed at the Stanford Nano Shared Facilities (SNSF), supported by the National Science Foundation under award ECCS-1542152. Use of the Stanford Synchrotron Radiation Lightsource, SLAC National Accelerator Laboratory, is supported by the U.S. Department of Energy, Office of Science, Office of Basic Energy Sciences under Contract No. DE-AC02-76SF00515. Access to MACS was provided by the Center for High Resolution Neutron Scattering, a partnership between the National Institute of Standards and Technology and the National Science Foundation under Agreement No. DMR-1508249. We acknowledge the support of the National Institute of Standards and Technology, U.S. Department of Commerce, in providing the neutron research facilities used in this work.

%\clearpage

\bibliography{HTO-cb}

\begin{thebibliography}{58}
\bibitem [{dis()}]{disc}%
  \BibitemOpen
  \href@noop {} {\bibinfo  {journal} {The identification of any commercial
  product or trade names does not imply endorsement or recommendation by the
  National Institute of Standards and Technology}\ }\BibitemShut {NoStop}%
\end{thebibliography}

\end{document}